\def\tit#1{``#1,''}
\long\def\comment#1{}

\let\ox=\otimes

\newcommand{\mx}[6]{{\left#1\begin{array}{rr}#2&#3\\#4&#5\end{array}\right#6}}
\newcommand{\Mx}[3]{\left#1\begin{array}{rrrr}#2\end{array}\right#3}

\newcommand{\ket}[1]{| #1 \rangle}

\newcommand{\brkt}[2]{\langle #1 \mid #2 \rangle}

\newcommand\U{\mathsf U}
\newcommand\St{\mathsf S}
\newcommand\Rt{\mathsf R}

\newcommand\eq[1]{Eq.~(\ref{#1})}

\documentclass{article}
\begin{document}
\begin{titlepage}
\def\newpage{\relax}
\def\abstractname{***}
\noindent{$_{}$\huge\hrulefill Extended Abstract}
\noindent\hrule
\title{\bf Classical programmability is enough for quantum circuits
universality in approximate sense}
\author{{\em Alexander Yu.\ Vlasov}\\
{\small FRC/IRH, 197101 Mira Street 8, St.--Petersburg, Russia}}
\date{}
\maketitle
\begin{abstract}
 It was shown by M. A. Nielsen and I. L Chuang \cite{Chuang97}, that it is
impossible to build {\em strictly} universal programmable quantum gate array,
that could perform any unitary operation {\em precisely} and it was suggested
to use probabilistic gate arrays instead. In present work is shown, that if
to use more physical and weak condition of universality (suggested already in
earliest work by D. Deutsch \cite{Deutsch85}) and to talk about
simulation with arbitrary, but finite precision, then it is possible to build
universal programmable gate array. But now the same no-go theorem by Nielsen
and Chuang \cite{Chuang97} will have new interesting consequence --- controlling
programs for the gate arrays can be considered as pure classical. More
detailed design of such deterministic quantum gate arrays universal
``in approximate sense'' is considered in the paper.
\end{abstract}
\end{titlepage}

\section{Introduction}
\label{sec:intro}

In the paper \cite{Chuang97} was discussed conception of programmable quantum
gate arrays, i.e., some quantum circuits are acting on a system in form
$\ket{d;P} \equiv \ket{d} \ox \ket{P}$ considered as {\em data register}
$\ket{d}$ and {\em program register} (or simply {\em program}) $\ket{P}$.
Similar with conception of usual classical computer, it was considered,
that circuit acts as some fixed unitary transformation $\U$ on whole
system and different transformations $U_P$ of data related only with content
$P$ of program register, i.e:
\begin{equation}
\U\bigl(\ket{d} \ox \ket{P}\bigr) = (U_P\ket{d}) \ox \ket{P'}.
\label{pga}
\end{equation}
It should be emphasized, here $U_P$ is same for any state of data register
$\ket d$ i.e., depends only on program $\ket P$, and states of these two
registers are not entangled before and after application of $\U$.

In relation with such a definition
in \cite{Chuang97} was noticed, that if to write the \eq{pga} for two
different programs $\ket{P}$ and $\ket{Q}$ and corresponding unitary data
transformations $\ket{U_P}$ and $\ket{U_Q}$, it is simple to find
by considering scalar products of both parts,
that if $U_P \ne \phi\,U_Q$ for some complex number $\phi$ then
the states of program register must be orthogonal for different programs:
\begin{equation}
\brkt{P}{Q} = 0.
\label{ortprog}
\end{equation}

The property \eq{ortprog} treated in \cite{Chuang97} as demonstration
that ``... no universal quantum gate array (of finite extent) can be realized.
More specifically, we show that every implementable unitary operation
requires an extra Hilbert space dimension in the program register.''
Due to such a problem in the article \cite{Chuang97} was suggested
idea of (exactly) universal stochastic programmable gate array developed
further by other authors \cite{Cirac2001}.

It was no-go theorem for ``universality in exact sense'' \cite{Cleve99}, but
fortunately already in initial definition of universal quantum computer
\cite{Deutsch85} was shown possibility of using some discrete everywhere
dense subset in the whole continuous space of unitary operators, it is called
sometime ``universality in approximate sense'' \cite{Cleve99}.

Here is shown, that in the approximate sense such universal programmable
quantum gate arrays with {\em possibility of approximation of any unitary
transformation of data register with given precision} are really exist and
constructions are quite simple and may be described directly. But the
property \eq{ortprog} is again very important, because if all different
possible states of program register are orthogonal, it is possible without
lost of generality to find implementation of same array with all different
programs are elements of computational basis and superpositions of the
states for the program register are {\em never used} and so any such array
can be designed with possibility of using only classical programs.
It is shown and discussed with more extent in relation with particular design
used below.

\section{Universal programmable quantum gate arrays}

It is more convenient here to use notation with program register first,
i.e., $\ket{P;d}$ instead of $\ket{d;P}$. The reason is more simple
form of operations like \mbox{\em CONTROL-$U$}:
\begin{equation}
 \Mx({1&0&0&0\\0&1&0&0\\0&0&u_{00}&u_{01}\\0&0&u_{10}&u_{11}}).
\end{equation}
Such operator acts on second qubit as $U$ only if first qubit is $\ket 1$,
but $\ket 0$ does not change anything.
A straightforward generalization for arbitrary $N \times N$ matrix $U$
and one program qubit is $2N \times 2N$ matrix:
\begin{equation}
 \mx(100U),
\label{CtlU}
\end{equation}
where 1 and 0 are $N \times N$ unit and zero matrices.

Let us now consider simple programmable gate array with dimension of
program is $M$ and data register is $N$. For quantum computation with
qubits $M = 2^m$ and $N = 2^n$. Let us introduce a matrix (cf. \cite{Jozsa95})
$\St = \St_{\{U_1,\ldots,U_M\}}$:
\begin{equation}
\St = \Mx({U_1&&&\mbox{\huge$0$}\\&U_2\\
                 &&\ddots\\\mbox{\huge$0$}&&&U_M}).
\label{qstep}
\end{equation}
where $U_1,\ldots,U_M$ are $M$ matrices $N \times N$. It is clear, that such
$NM \times NM$ matrix $\St = \St_{\{U_1,\ldots,U_M\}}$ acts as:
\begin{equation}
 \St\bigl(\ket{P} \ox \ket{d}\bigr) =
 \ket{P} \ox (U_P\ket{d}).
\end{equation}
and it corresponds to \eq{pga} with $P = P'$ is simply number of matrix $U_P$.

Such matrix was suggested for description of conditional quantum dynamics
in \cite{Jozsa95}, but \eq{ortprog} introduced in \cite{Chuang97} and
discussed here shows also, that we do not need more general control with
superpositions of basis states of program (control register).

Let us consider a case with $\{U_1,\ldots,U_M\}$ are universal set of
gates (in approximate sense) for some quantum circuits with $n$ qubits
together with unit matrix $U_0 \equiv \mathbf 1$. Here $N=2^n$ and for design
with matrix $\St$ size of program is small enough --- number of
qubits in the register is $m=\lceil\log_2 M \rceil$. But if due to
technical problem with implementation we should use set of {\em CONTROL-$U_i$}
gates \eq{CtlU} instead of $\St$ \eq{qstep}, then number of program qubits
are $m=M$.

We may use small universal sets with only one and two-qubit gates
\cite{Cleve99,Gates95,DiVincenzo95,Vlasov2000}. For example,
if to choose a set with $n+2$ one-qubit gates and $n-1$ two-gates
acting on pairs of neighboring qubits from \cite{Vlasov2000} together
with unit matrix, then number of qubit in control register is
$m=\lceil\log_2(2n+2)\rceil$ in best case and even in worst case discussed
earlier it is $2n+2$.

Let us now extend program register up to $k m$ qubits, i.e. state
of arrays described now as $\ket{P_k;\ldots;P_2;P_1;d}$
and together with matrix $\St$ acting only on last $m+n$ qubits
$\ket{P_1;d}$ as described above let us consider unitary matrix $\Rt$
acting on program register as {\em right cyclic shift}:
\begin{equation}
 \Rt\,\ket{P_k;\ldots;P_2;P_1;d} = \ket{P_1;P_k;\ldots;P_2;d}
\end{equation}

It is clear, that if $\U = \Rt \St$, then operator $\U^k$ can perform
with data arbitrary sequence of up to $k$ operators $\{U_1,\ldots,U_M\}$:
\begin{equation}
 \U^k\bigl(\ket{0;\ldots;0;P_l;\ldots;P_1}\ox\ket{d}\bigr) =
 \ket{0;\ldots;0;P_l;\ldots;P_1}\ox(U_{P_l}\cdots U_{P_1}\ket{d})
\end{equation}
here $U_0$ is unit operator used to fill out $k-l$ positions for $l < k$
and $U_{P_l}\cdots U_{P_1} \cong U$ is arbitrary product of gates
from the finite universal set $\{U_1,\ldots,U_M\}$ for approximation of
any data gate $U$ with necessary precision. Usually number of terms
in product $l \gg M$.
If minimal error of simulation is given, it is possible to find necessary
number $k$ and then $\U^k$ is universal circuit with bounded chosen accuracy.

The idle $k-l$ steps make more clear advantage of a design well known in
classical theory of computations --- instead of consideration of whole
circuit $\U^k$ it is possible to introduce notion of {\em one computational
step} $\U$. In such approach action of quantum circuit similar with usual
CPU ``timing'' with repeating of same $\U$ till halting.

It is also useful, then due to algorithm of approximation we need to apply
same loop many times, for example instead of some operator $U$ may be used
``$j$-th root'' $U^{1/j}$ \cite{Ekert95,Vlasov2000} and the circuit
should be applied $j$ times and can be expressed as $\U^{k\,j}$.

Here were examples of some general architecture for programmable quantum gate
arrays --- we have $n$ data qubits and program with $p+m$ qubits of two
different kinds. Here are $m$ control qubits of quantum controller
together with $n$ data qubits form input for ``quantum step operator''
$\St$ described by \eq{qstep}.

Generalization of $\Rt$ operator is not necessary simple rotation, because
such model may use too many space. The $\Rt$ is equivalent of reversible
classical\footnote{``Qu-ERCC'' --- Quantum Equivalent of Reversible Classical
Circuit, or simply ``classical gate'' in terminology used by \cite{Ekert95}}
circuit acting on $p+m$ (qu)bit program register with purpose to generate
necessary index of universal quantum gate for quantum controller during
simulation.

It should be mentioned also yet another advantage and special property
of ``pseudo-classical'', i.e., orthogonal states of program register. Only
for such kind of programs it is possible to make measurements of state
without destruction and so use tools like {\em halt} bit. It is also
further justification of notion of {\em variable lenght} algorithms already
used above.

\medskip

The consideration shows possible model of programmable ``quantum chip''.
It has three different kinds of ``wiring'': quantum, intermediate and
classical ``buses'' with $d$, $m$ and $p$ (qu)bits respectively and two
different kinds of circuits: quantum controller acting on quantum bus and
controlled by intermediate bus and reversible classical (or Qu-ERCC) circuit
with input via classical bus for control of intermediate bus.

Due to result about orthogonality of programs \cite{Chuang97} the design
is general for deterministic circuits with pure states, and does not
need some special quantum control, i.e, superpositions in program register,
only data is necessary quantum.

On the other hand, for circuits with mixed states, stochastic programmable
quantum arrays \cite{Chuang97,Cirac2001} control already is not limited by
the (pseudo) classical case discussed here.

\end{document}